\documentclass[twocolumn]{aastex6}
\RequirePackage{lineno}
\usepackage{amsmath}
\usepackage{amssymb}
\usepackage{amsthm}
\usepackage{natbib,url,bm}
\usepackage{array}
\usepackage{float}
\usepackage{graphicx}
\usepackage{subfigure}
\usepackage{color}

\newcounter{ichi}
\newcounter{ni}
\newcounter{san}
\newcounter{yon}
\def\be{\begin{equation}}
\def\ee{\end{equation}}
\def\ba{\begin{eqnarray}}
\def\ea{\end{eqnarray}}

\usepackage{ulem,color}
\usepackage[utf8]{inputenc}

\slugcomment{draft v1}

\shorttitle{Afterglow of the fast tail of  merger ejecta}
\shortauthors{Hotokezaka et al.}

\begin{document}


\title{
Synchrotron radiation from the fast tail of dynamical ejecta of neutron star mergers}


\author{Kenta Hotokezaka\altaffilmark{1}}
\author{Kenta Kiuchi\altaffilmark{2}}
\author{Masaru Shibata\altaffilmark{2,3}}
\author{Ehud Nakar\altaffilmark{4}}
\author{Tsvi Piran\altaffilmark{5}}


\altaffiltext{1}{Department of Astrophysical Sciences, Princeton University, Peyton Hall, Princeton, NJ 08544, USA}
\altaffiltext{2}{Center for Gravitational Physics, Yukawa Institute for Theoretical
Physics, Kyoto University, Kyoto, 606-8502, Japan}
\altaffiltext{3}{Max Planck Institute for Gravitational Physics (Albert Einstein
Institute),
Am Mühlenberg 1, Postdam-Golm 14476, Germany}
\altaffiltext{4}{The Raymond and Beverly Sackler School of Physics and Astronomy, Tel Aviv University, Tel Aviv 69978, Israel}
\altaffiltext{5}{Racah Institute of Physics, Hebrew University, Jerusalem, 91904, Israel}


\begin{abstract}
We find, using high resolution numerical relativistic  simulations, that the  tail of the dynamical ejecta of neutron star  mergers extends to  mildly relativistic velocities faster than $0.7c$. The  kinetic energy of this fast tail is $\sim 10^{47}$--$10^{49}$ erg, depending on the neutron star equation of state and on the binary masses.  
The synchrotron flare arising from the interaction of this fast tail  with the surrounding ISM can power the observed  non-thermal emission that followed  GW170817, provided that the  ISM density is 
$\sim 10^{-2}\,{\rm cm^{-3}}$, the two neutron stars had roughly equal masses and  the neutron star equation of state is soft (small neutron star radii).
One of the generic  predictions of this scenario is that the cooling frequency crosses the X-ray band on a  time scale of a few months to a year, leading to a cooling break in the X-ray light curve. 
If this dynamical ejecta scenario is correct, we expect that the synchrotron radio flare  from the ejecta that have produced the macronova/kilonova emission will
be observable on time scales of $10^3$ to $10^5$ days.
Further multi-frequency observations will confirm or rule out this dynamical ejecta scenario.

\end{abstract}


\keywords{
supernovae: general --- 
pulsars: general  --- 
binaries : close --- 
}

\section{Introduction}
The discovery of the gravitational waves from 
a neutron star merger, GW170817,
has opened a new era of multi-messenger astronomy \citep{GW170817, Abbott2017ApJ}.
The gravitational-wave signal  was followed by multi-frequency electromagnetic emission 
including a $\gamma$-ray pulse, uv/optical/IR macronova/kilonova, 
and long lasting non-thermal emission ranging from radio to X-rays.
The macronova/kilonova observations show that 
the typical  velocity of the ejecta is $\sim 0.1$--$0.3c$  and
$r$-process elements of a mass of 
$\sim 0.05\,M_{\odot}$ are synthesized in the ejecta if the radioactive decay powers the emission
(e.g., \citealt{Andreoni2017,Arcavi2017Natur,Cowperthwaite2017,Evans2017,Lipunov2017,Kasliwal2017,Kilpatrick2017,Smartt2017,Valenti2017,Drout2017,Pian2017,Swope,Tanvir2017,Utsumi2017}, and see also the modelings, e.g.,
\citealt{Kasen2017, Tanaka2017, Waxman2017, Perego2017,Villar2017}). 
This mass 
estimate together with the rate of GW170817 supports early predictions \citep{Lattimer1974,eichler1989Nature}  that 
$r$-process elements in our Galaxy are predominately produced
by mergers (e.g. \citealt{Cote2017,rosswog2017,hotokezaka2018}). However, the mechanism that ejects
such a large amount of material still remains an open question (e.g. \citealt{Metzger2018,Shibata2017}).

X-ray and radio signals were discovered at 9 and 16 days after the merger, respectively
\citep{Haggard2017, Troja2017, Hallinan2017}.
These signals are most likely explained by synchrotron radiation 
arising from the shock formed between the merger outflow and interstellar
medium (ISM). Both off-axis and on-axis emission models are consistent with the observed data
up to $\sim 30$ days  \cite[e.g.,][]{Hallinan2017,Troja2017,Margutti2017, Gottlieb+17B}. 
However,  radio observations up to $\sim 100$ days 
show that the flux density  continued to rise  as $\sim t^{0.8}$ \citep{Mooley2017}.
 X-ray and optical observations subsequently confirmed this behavior
 \citep{Ruan2017,Lyman2018,Margutti2018}.

{Recently, \cite{Nakar2018} have shown that
this moderately rising light curve  implies  that the emitting matter is  moving towards us at the time of the emission, 
namely, the matter   is moving within an angle $\theta <1 /\Gamma$ ($\Gamma$ is the Lorentz factor of the matter) from the line of sight towards us, otherwise the light curve would have risen much faster ($F_\nu \propto t^\alpha$
with $\alpha > 3$). They have also shown a continuous energy injection into the blast wave.  The isotropic equivalent energy increases like $E_{\rm iso}(>\Gamma) \propto  
 \Gamma^{-{8\alpha+6(p-1)}/({3-\alpha})}$ (where $p$ is the electron's spectral index),
otherwise the light curve would have declined. }

{This energy injection implies that the outflow must have a structure, either radial or angular or both. Such a structure can arise naturally  from  the interaction of the jet with the surrounding ejecta that forms a cocoon \citep[see e.g.][]{Gottlieb2018MNRAS,Kasliwal2017,Gottlieb+17B,Lazzati2017,Nakar2018}. 
This cocoon can arise in the case that the jet is choked or  it emerged and generated a GRB pointing elsewhere. Note that a successful jet and  cocoon are  sometimes called in the literature  ``a structured jet" \citep[see e.g.][]{Margutti2018}.
}

Here, we explore a  third possibility in which the  observed synchrotron emission arises due to the fast tail of the  dynamical ejecta and it has nothing to do with   the question whether a jet existed or not  and how it evolved. 
In this dynamical ejecta scenario, suggested long ago by \cite{NP11} and elaborated  by 
\cite{piran2013MNRAS,HP2015}  and others,  the synchrotron emission arises from the interaction of 
the mergers' dynamical ejecta with the surrounding ISM. These earlier studies focused on the late time emission that would arise from the sub-relativistic component of the outflow, but stressed that a very strong early signal is expected if a faster component exists. 

\begin{figure*}[t]
\includegraphics[scale=0.7]{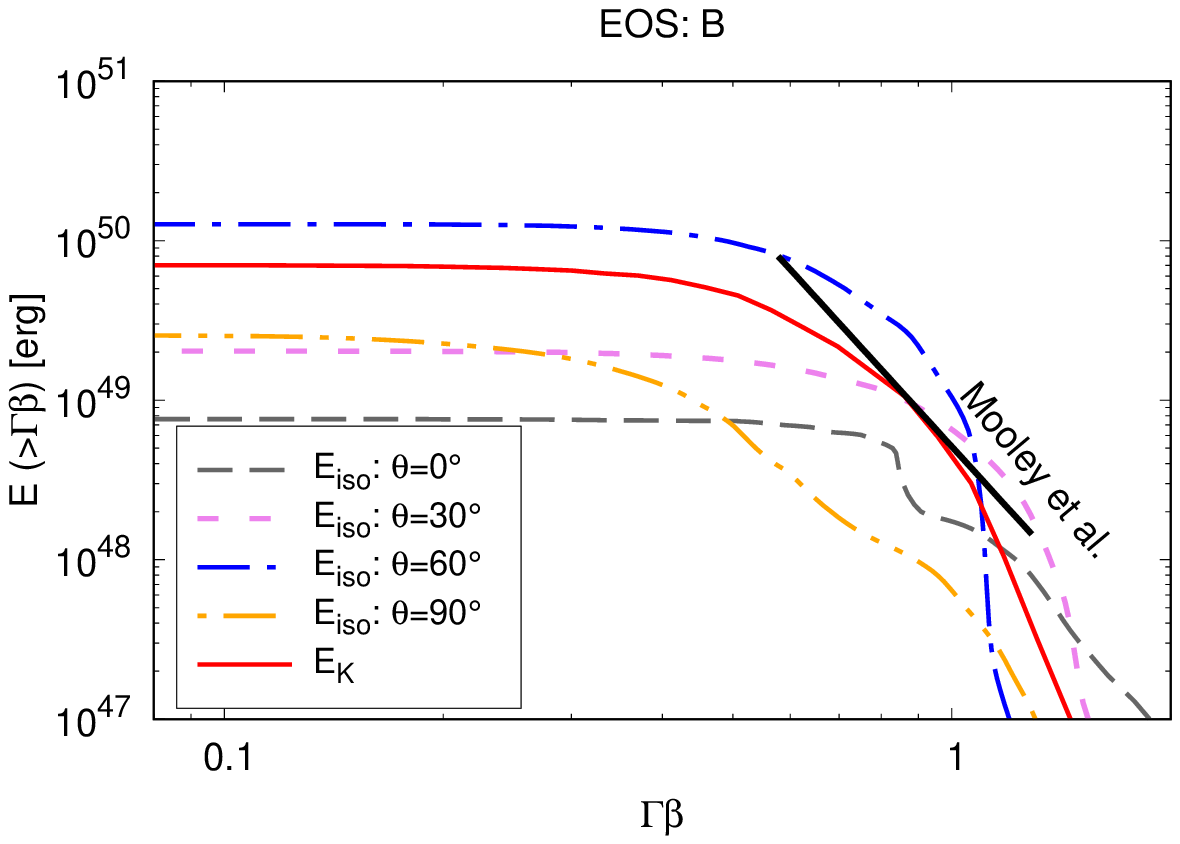}
\includegraphics[scale=0.7]{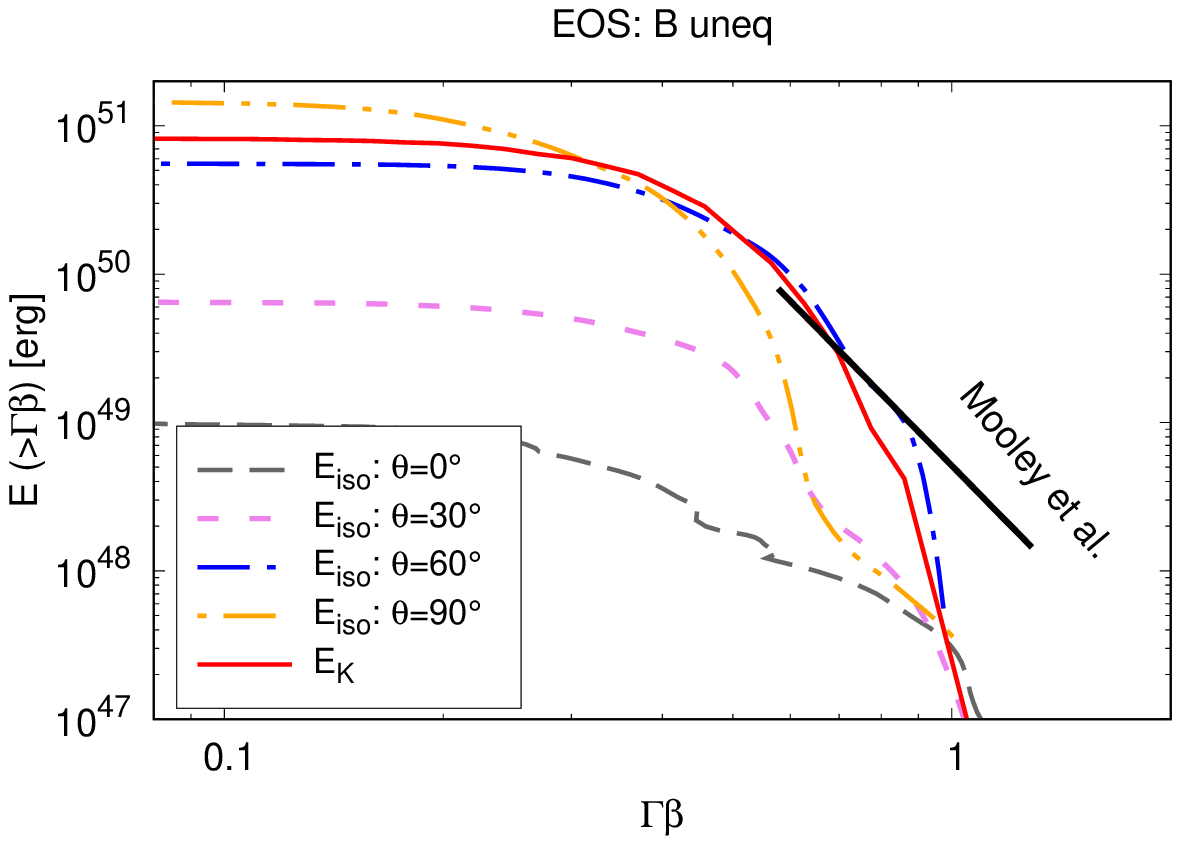}
\caption{
The total kinetic energy and the isotropic equivalent kinetic energy distributions of the dynamical merger ejecta
at different polar angles for models B ({\it left}) and
B uneq ({\it right}). Also shown 
as a solid straight line is the slow quasi-spherical model of \cite{Mooley2017}.
}
\label{fig:E}
\end{figure*}

\begin{figure*}[t]
\includegraphics[scale=0.7]{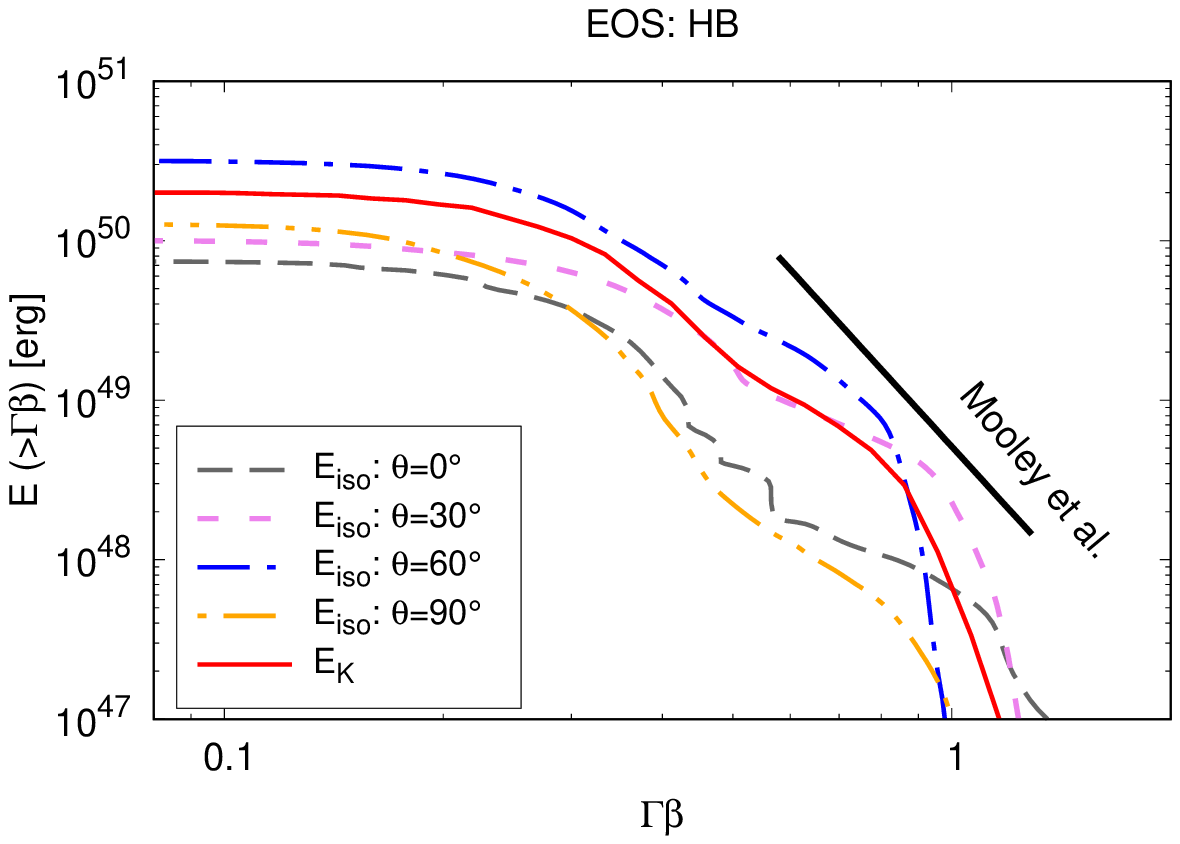}
\includegraphics[scale=0.7]{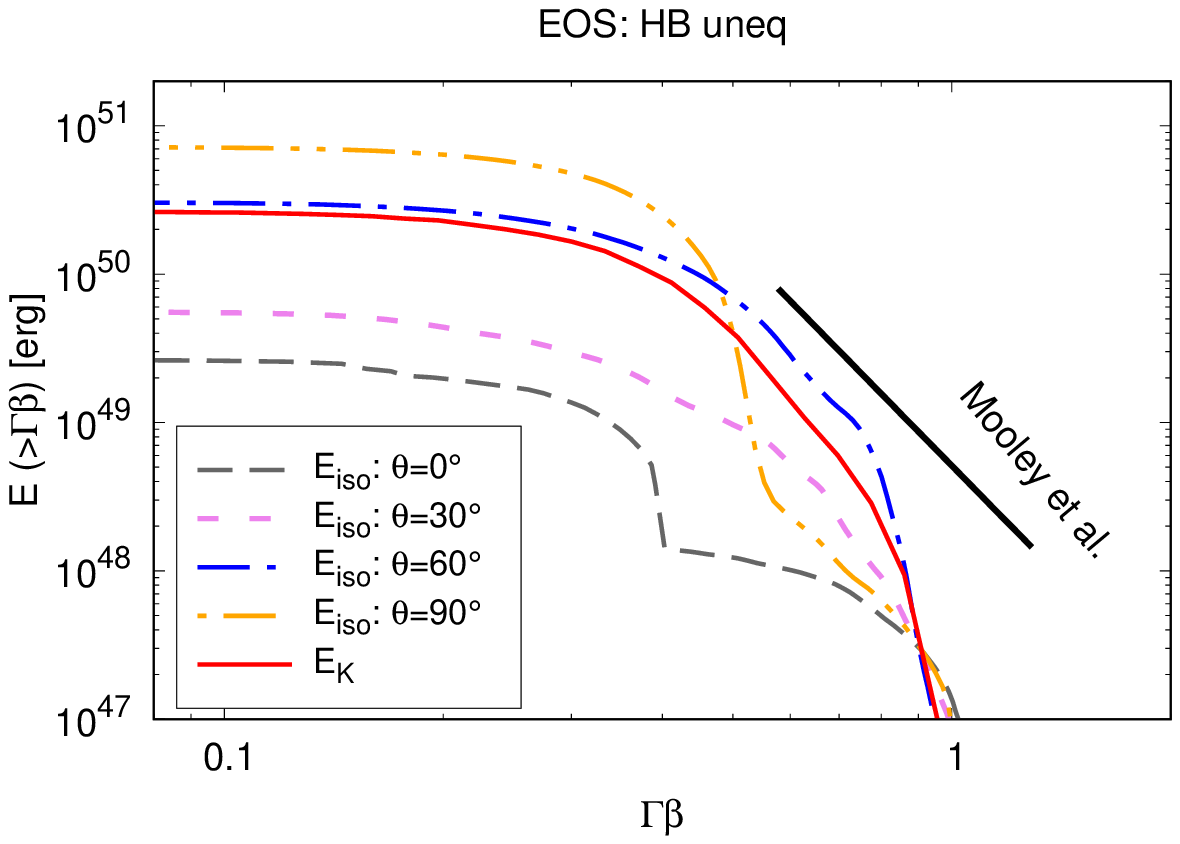}
\caption{
Same as figure \ref{fig:E} but for model HB. 
}
\label{fig:E3}
\end{figure*}

Indeed, 
a high velocity (mildly relativistic) tail of the ejecta can also explain the observed emission that followed   GW170817   \citep{Mooley2017}. 
Such a high velocity component was expected 
but it was very difficult to estimate since only a very small amount of matter moves at these high velocities.
The profile of the high velocity tail was poorly known because of the complexity 
of hydrodynamics of  mergers.
Previous attempt to calculate this  involved both analytic considerations \citep{Kyutoku2014}
and numerical simulations \citep{hotokezaka13b,bauswein2013ApJ}.  The latter were limited by their resolution.  Here we use the results of a new series of numerical simulation \citep{Kiuchi2017} with a much higher spatial resolution that allows us to explore in details the velocity profile. These simulations find, in all cases considered,  a light fast component with very steep energy profiles (see figures \ref{fig:E} -- \ref{fig:E2}). 
It is remarkable that \cite{Gottlieb+17B}, who considered the conditions  for a cocoon shock breakout to produce the observed $\gamma$-rays, found that  a high velocity component with a comparable amount of matter and energy profile is needed to explain these observations.

The structure of the paper is as follows: we show in \S\ref{sec:ejecta} that  recent high-resolution numerical relativity simulations reveal such  high velocity components and discuss their properties. 
In \S\ref{sec:radio} we use these profiles to  calculate the synchrotron radiation light curves arising from the
shock between the ejecta and ISM and compare them with the observed data. Finally, in \S\ref{sec:discussion}, we discuss the generic features 
and future expectations of  the dynamical ejecta scenario and summarize our conclusions in \S\ref{sec:conclusions}.

\section{The fast tail of dynamical ejecta}
\label{sec:ejecta}

A small fraction of the dynamical merger ejecta accelerates to 
mildly relativistic velocities when the shock formed between 
the colliding neutron stars emerges from the surface \citep{Kyutoku2014}.
This component is the high velocity tail of the dynamical ejecta.
The flux of the synchrotron radiation arising from the ISM-ejecta shock can be 
significantly enhanced by even  a small amount of a fast ejecta
\citep{piran2013MNRAS,HP2015}. 
While fast ejecta  have been found in previous numerical merger simulations
\citep{hotokezaka13b,bauswein2013ApJ,Metzger2015},
its amount was not clear   due to the low numerical resolution of those earlier simulations.

We begin by describing  the ejecta profile obtained from recent high resolution 
numerical relativity simulations  \citep{Kiuchi2017},
whose the finest grid spacing, $\Delta x$, satisfies: $\Delta x\sim 70$ m.
We discuss five  models  with a total mass
of $\approx 2.7M_{\odot}$, a mass ratio of $1$ or $\approx 0.8$,
and three different neutron star equations of state: models B, HB, and H
\citep{Kiuchi2017}. 
These models are consistent with the mass and tidal deformability 
constraints of GW170817 \citep{GW170817}.
The parameters of these models {and the resulting total kinetic energy and mass of the ejecta }  are listed in Table. \ref{tab:model}.  Hereafter, we refer  to equations of state  which give 
small (large) radius neutron stars as 
``soft" (``stiff") equations of state.

{Note that, like in practically all other numerical relativistic merger simulations \citep{hotokezaka13b,bauswein2013ApJ,sekiguchi2015PRD,radice2016MNRAS,Dietrich2017,Shibata2017} in all models the total mass 
of dynamical ejecta is lower than the one needed to power the observed macronova/kilonova. 
This additional mass can be driven by 
other  mass ejection mechanisms, such as winds from the surrounding accretion disk  or from the hypermassive neutron star  that have not been included in these merger simulations (see, e.g., \citealt{dessart09,fernandez2013MNRAS,metzger2014MNRAS,perego2014MNRAS,just2015MNRAS,fernandez2015MNRAS,fujibayashi2017,Siegel2017}). However,  the typical velocity of this wind ejecta is 
much slower than that of the dynamical ejecta, and hence, this ejecta is not relevant for the early synchrotron
emission focused in this paper.} 

\begin{figure}
\includegraphics[scale=0.7]{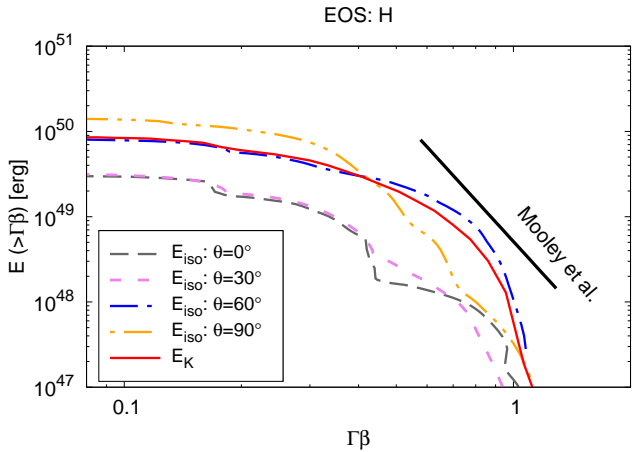}
\caption{Same as figure \ref{fig:E} but for model H with equal masses.}
\label{fig:E2}
\end{figure}

Figures \ref{fig:E} -- \ref{fig:E2} depict the kinetic energy distributions as a function of
$\Gamma \beta$ 
obtained from the simulations. 
Here,  $\Gamma$ is the Lorentz factor and $\beta$ is the velocity in units 
of the speed of light.
The total  kinetic energy of the ejecta is  in the range from 
$\sim 7\cdot10^{49}$ to $8\cdot 10^{50}$ erg. 
The ejecta of the softest model B is faster than that of models HB and H.
This can be explained if the shock formed at the collision is 
stronger for more compact neutron stars, and thus, the ejected
material is faster.  For unequal mass cases the ejecta are predominantly produced  
through the tidal disruption of the lower mass neutron star and it contains 
a larger amount of slow material and a smaller amount of fast material compared
to the equal mass cases. 
{ Note also that the ejecta mass of model B is much smaller
than the other models because the merging neutron stars
promptly collapse into a black hole, and thus, the mass ejection occurs 
on only a  short time scale.}

\begin{table}[t]
\begin{center}
\caption[]{Parameters }
\label{tab:model}
\begin{tabular}{lcccc}
\hline \hline
Model & $m_1,\,m_2$ [$M_{\odot}$]  & $R_{\rm ns}\,[{\rm km}]$ & $M_{\rm ej}\,[M_{\odot}]$  & $E_{\rm ej}$\,[erg]    \\ \hline
B & $1.35,\,1.35$ & $10.96$ & $5\cdot10^{-4}$ & $7\cdot 10^{49}$ \\
HB & $1.35,\,1.35$ & $11.61$ & $3\cdot 10^{-3}$ & $10^{50}$ \\
H & $1.35,\,1.35$ & $12.27$ & $3\cdot10^{-3}$ & $9\cdot10^{49}$ \\ \hline
B uneq& $1.51,\,1.21$ & $10.96$ & $10^{-2}$ & $8\cdot 10^{50}$ \\
HB uneq& $1.51,\,1.21$ & $11.61$ & $5\cdot 10^{-3}$ & $3\cdot 10^{50}$ \\
\hline \hline 
\end{tabular}
{\footnotesize{ $R_{\rm ns}$ is the radius of a non-rotating cold  neutron star of $1.35M_{\odot}$.} 
}
\end{center}
\end{table}

The isotropic equivalent energy distributions at different polar angles
are also shown in Figs. \ref{fig:E} -- \ref{fig:E2}, demonstrating the anisotropy of the ejecta.
For the equal mass cases, the material ejected at $30^{\circ}$--$60^{\circ}$
has more kinetic energy.  For the unequal mass cases,
more material is ejected on the equatorial plane. 
Another important feature is that the ejecta
moving faster than $\Gamma \beta \sim 1$ is somewhat isotropic for all the cases.
{Therefore, the early light curve arising from this mildly relativistic tail is 
expected to be rather isotropic. }

Also shown in figures \ref{fig:E}--\ref{fig:E2}, for a comparison, is the
quasi-spherical slow ejecta model used by  \cite{Mooley2017}:
\begin{eqnarray}
E(>\Gamma \beta)=5\cdot 10^{50}\,{\rm erg\,} (\Gamma \beta/0.4)^{-5}~~~({\rm for}\,\beta <0.8)\ . \label{kunal}
\end{eqnarray}
This distribution can fit the observed data up to $\sim 100$ days after GW170817. 
 Note that the non-thermal afterglow up to $\sim 100$ days is produced by a component faster than 
 $\beta \approx 0.6$ \citep{Mooley2017} and,
 of course, this profile cannot be extended to lower velocities $\beta \ll 0.4$.
 The total kinetic energy distributions for model B is the closest to this distribution. 

\section{The  dynamical ejecta emission}
\label{sec:radio}
The interaction between the expanding merger ejecta and the surrounding ISM results in 
a shock, in which particle acceleration and magnetic field amplification 
occur. As a result, this blast wave emits multi-wavelength
synchrotron radiation  \citep{NP11}.
Here, we numerically calculate the synchrotron
radiation emitted by the accelerated electrons since we are interested in
mildly relativistic blast waves, for which either limits, $\beta\ll1$,
or $\Gamma\gg1$, cannot be used.
To follow the evolution of the ejecta expanding in the ISM with a constant density, $n$, 
we solve the adiabatic radial expansion of the ejecta at each polar angle
using the kinetic energy distributions given in the previous section.
The synchrotron flux from the blast wave element at each solid angle
 at a given observer time  is  calculated assuming
a power-law distribution of electrons, $dN/d\gamma_e\propto \gamma_e^{-p}$,
and the standard equipartition parameters $\epsilon_e$ and $\epsilon_B$,
which reflect the conversion efficiency of the internal energy of the shocked ISM to the
accelerated electrons' energy and magnetic energy respectively (e.g. \citealt{SPN98,NP11} and see \citealt{HP2015} for details). We fix the viewing angle to be $30^{\circ}$ \citep{Abbott2017ApJ} and set $\epsilon_e$ to be $0.1$.
Therefore, there are in total  three free parameters, $n$,  $\epsilon_B$, and $p$, with which we fit the observed data in  our afterglow modeling.

\begin{figure*}
\includegraphics[scale=0.7]{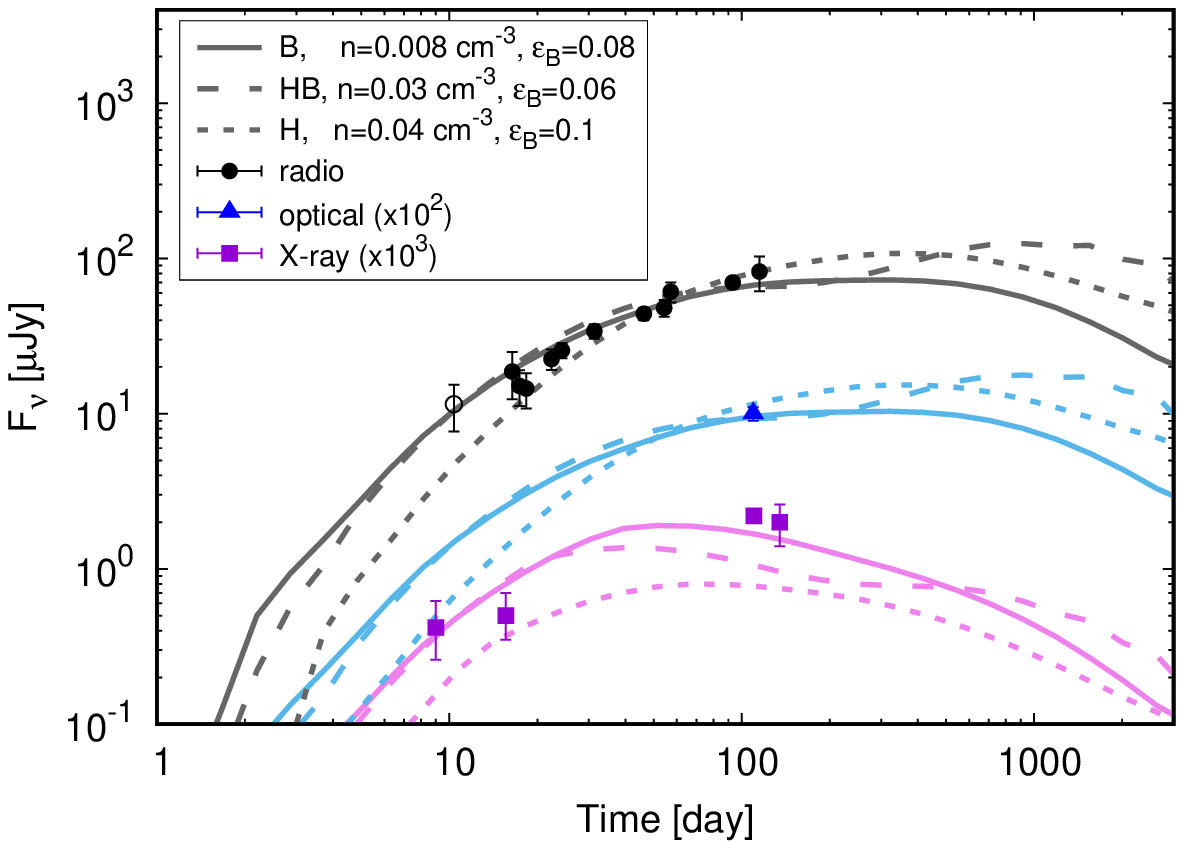}
\includegraphics[scale=0.7]{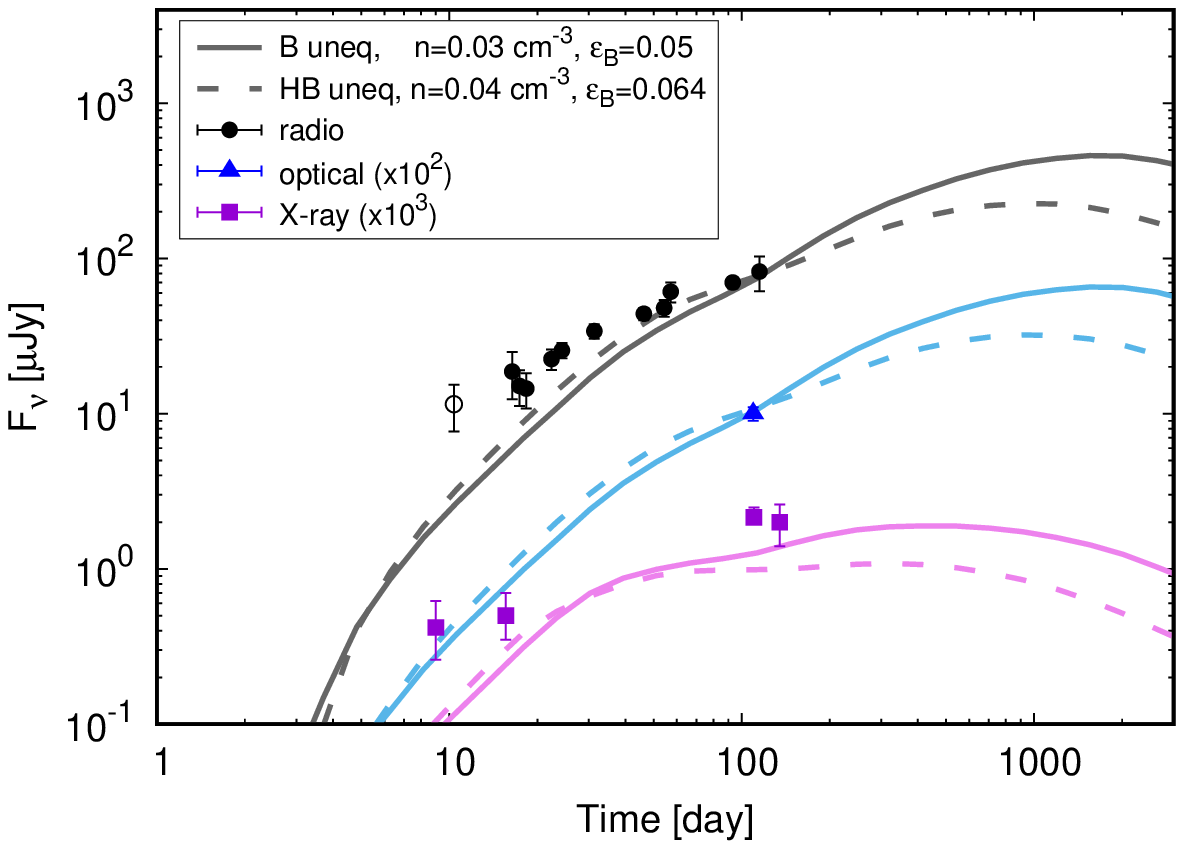}
\caption{
Light curve models of the synchrotron flare that followed  GW170817. {\it Left} and {\it right} panels
show the light curves for the equal mass and unequal mass cases, respectively. 
The solid circles, triangle, and squares
show the radio data ($3$\,GHz; \citealt{Hallinan2017,Mooley2017,Margutti2018}), optical data (r-band; \citealt{Margutti2018,Lyman2018}), and X-ray data (1 keV; \citealt{Haggard2017,Troja2017,Margutti2017,Margutti2018, Ruan2017, D'Avanzo2018,Troja2018}). 
The optical  and X-ray flux densities are  multiplied by  factor of $10^{2}$  and $10^{3}$ correspondingly.
The open circle shows a marginal
detection at $6$\,GHz \citep{Hallinan2017} corrected to the flux density at $3$ GHz assuming a spectral index of $-0.55$.
The solid, dashed, dotted lines show the theoretical light curve arising from 
the dynamical merger ejecta for model B, HB, and H, respectively. Here we assume microphysics parameters, $\epsilon_e=0.1$ and $p=2.1$, and a viewing angle of $30^{\circ}$.
}
\label{fig:L}
\end{figure*}

Figure \ref{fig:L} shows the calculated light curves for the different models as well as the observations  of GW170817. 
The light curves match the observed data, 
for 
$n=0.008$--$0.04\,{\rm cm^{-3}}$,  $\epsilon_B= 0.05$--$0.1$,
and $p= 2.1$.  
Note that even the higher values of the ISM density are  consistent with 
the upper limit on the mean ISM density of NGC 4993\footnote{This upper limit is for a neutral hydrogen component of the ISM.
 The mean density of hot ionized gas around GW170817 is currently not constrained.}, $n\lesssim 0.04\,{\rm cm^{-3}}$,  inferred from
the upper limit on the HI mass  \citep{Hallinan2017}.
For the equal mass cases,  the radio and optical light curves 
continuously rise up to $\sim 100$ days  and then have a plateau phase lasting a few hundred days. 
This early component 
is produced by the tail of the ejecta faster than $\Gamma \beta \sim 0.6$.
Model H  contains a  lower  
kinetic energy at $\Gamma \beta \sim 1$. As this  fastest  component dominates the early light curve the flux density up to $20$ days is lower than those of model B and HB.

For unequal masses, the flux densities continuously rise until $1000$ days because
of the larger amount of material at low velocities. However, the slope of the calculated 
light curves is steeper than the observed one because the kinetic energy distribution
of the these  models declines more steeply
 at high velocities.

An important feature of the X-ray light curves 
is that they peak on a time scale of a few months to a year and then their temporal evolution is different from that of the radio and optical light curves once   the synchrotron cooling 
frequency becomes lower than $1$ keV.  
For model H, in particular,  the cooling frequency falls  below 1 keV  
at early times $\sim 10$ days because of the required relatively large values of the ISM density and $\epsilon_B$.
Note that the exact location of the cooling frequency can be
higher due to several uncertainties (see the discussion in the next section). However,
it is generally expected that the cooling frequency crosses the X-ray band
on a time scale of a few months to a year in this  scenario. 


%

\section{Discussion}
\label{sec:discussion}
{\it Cooling frequency:} 
An important feature of the synchrotron light curves obtained 
in \S \ref{sec:radio} is the evolution of the synchrotron cooling frequency,
that crosses the X-ray band on a time scale of a few months to a year.  After this time, 
the  X-ray flux density declines with time faster than the radio and optical flux densities.
In the Newtonian limit ($\beta\ll 1$), the cooling frequency at a given time, $t$, 
is approximately estimated as (e.g., \citealt{SPN98})
\begin{eqnarray}
\nu_c & \sim &0.5\, {\rm keV}  \left(\frac{\beta}{0.6}\right)^{-3} \left(\frac{n}{0.01\,{\rm cm^{-3}}}\right)^{-3/2} \\
& & ~~~~~~~~~~~~~~\times\left(\frac{\epsilon_{B}}{0.05} \right)^{-3/2} \nonumber
\left(\frac{t}{100\,{\rm days}} \right)^{-2},\label{eq:nuc} 
\end{eqnarray}
or the cooling frequency for a given flux density is
\begin{eqnarray}
\nu_c  \propto \beta^{\frac{5p-6}{3}}n^{\frac{p-4}{6}}\epsilon_B^{\frac{p-8}{6}}
\epsilon_e^{\frac{2(p+1)}{3}}F_{\nu}^{-\frac{2}{3}},
\end{eqnarray}
where $F_{\nu}$ is the flux density at a frequency in the range of $\nu_a,\,\nu_m<\nu<\nu_c$. 
Note that, for a given flux density, the cooling frequency  increases by 
decreasing $n$ and $\epsilon_B$ and by increasing the velocity for the expected range of 
the electrons' power-law index, $2<p<3$.
In other words, it depends sensitively on the kinetic energy of the fast tail.
The cooling frequency  can be higher than those of our models if 
the kinetic energy at velocities faster than $\beta\sim 0.6$ is only slightly larger.
For instance, the kinetic energy distribution of Eq. (\ref{kunal}) results in $\nu_c \gtrsim 10$ keV at $100$ days \citep{Mooley2017}. 
Note also that the cooling frequency is much higher for the cocoon  models \citep{Mooley2017,Nakar2018,Margutti2018,Lyman2018}, that involve an outflow much faster  than the fast tail of the dynamical ejecta \citep[see, e.g.,][]{Gottlieb+17B,Lazzati2017,Nakar2018}.

The X-ray observations around $100$ days
show that the X-ray flux density and the photon index in the X-ray band  are
consistent with a single power-law spectrum from the radio to  the X-ray bands \citep{Mooley2017, Margutti2018, Ruan2017, D'Avanzo2018,Troja2018}. 
This suggests that the cooling frequency is above the X-ray band, ruling out models with much lower   cooling frequency, e.g., model H. 
However, 
it should be noted that 
the estimate of equation (\ref{eq:nuc}) and  the cooling frequency used in our light curve modeling are correct only within 
an order of magnitude.  In fact, \cite{GranotSari02} obtained the spectral breaks of the afterglows
for the Blandford-McKee solution and found that  the cooling frequency was higher by a factor of $\gtrsim 3$  than  the simple order of magnitude estimate \citep{SPN98}. 
Therefore, 
 models B and HB for the equal mass caes can be considered as  consistent with the current observed data.  
However, a generic expectation of the dynamical ejecta scenario
 is  that the cooling frequency will cross the X-ray band on a time scale not much longer than a few months. 
 Therefore further multi-frequency observations  will easily confirm or rule out this model.

{\it Late-time signal:} 
In addition to the fast moving  dynamical ejecta,  there should be a main, sub-relativistic   ejecta component
that have produced the uv/optical/IR macronova/kilonova signal. As already mentioned 
the  dynamical ejecta masses of our models,
$5\cdot 10^{-4}$--$10^{-2}M_{\odot}$,
are much smaller than those estimated from the macronova observations, 
$\sim 0.05M_{\odot}$  (e.g., \citealt{Kasliwal2017}).
This component has the photospheric velocities of $\sim 0.1$--$0.3c$ and a kinetic energy of $\sim 10^{51}$\,erg,
which is also larger than the kinetic energy of the dynamical ejecta  in our dynamical ejecta models\footnote{The typical velocity of the macronova ejecta can be slower than
the photospheric velocities of $0.1$--$0.3c$. Numerical simulations show
that the velocity of the wind outflows, that we consider here as the dominant 
outflow component producing the macronova/kilonova emission, 
 is typically $\sim 0.05$--$0.1c$
(e.g., \citealt{fernandez2013MNRAS,fujibayashi2017,Siegel2017}). }. 
Some other processes, not taken into account in the simulations used in this work, 
must be responsible for the ejection of this additional mass.

\begin{figure}[t]
\includegraphics[scale=0.7]{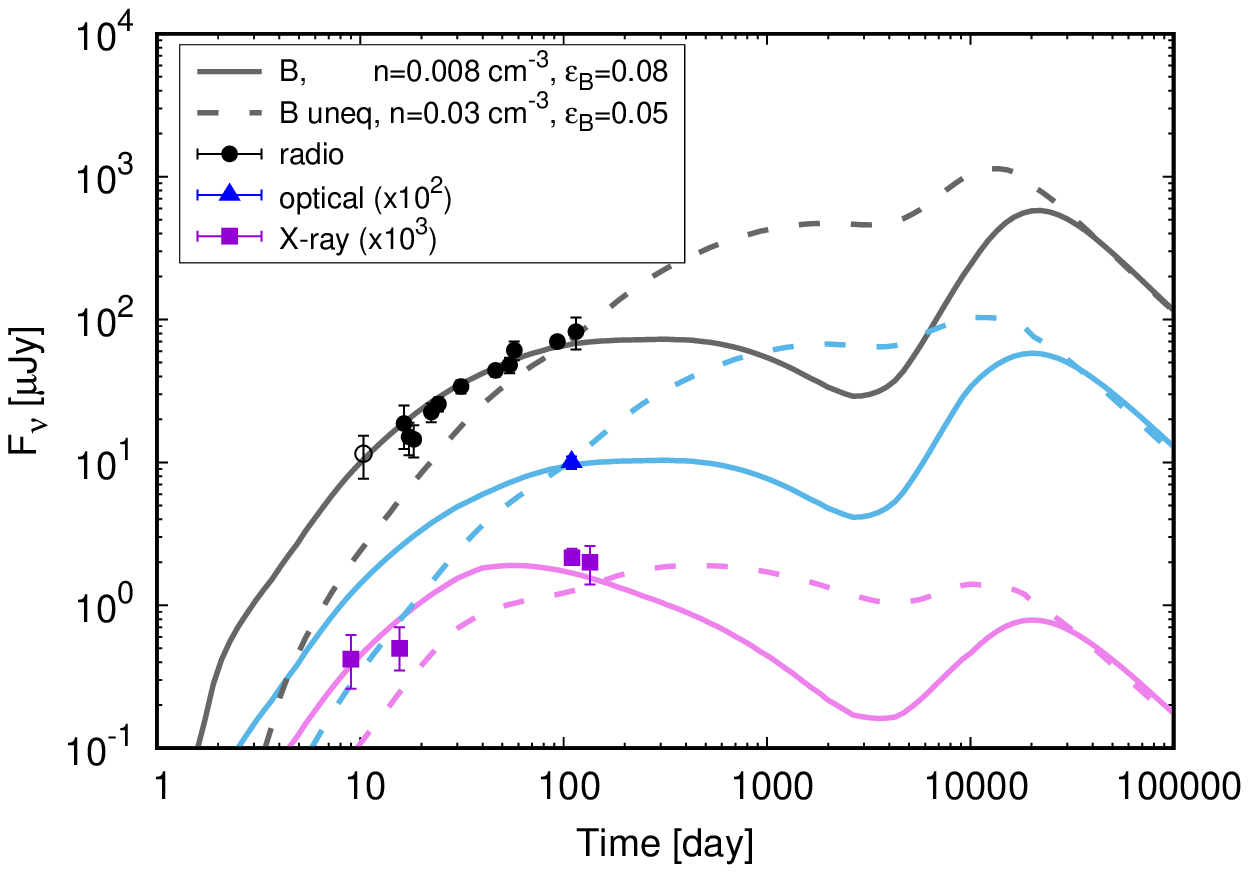}\\
\includegraphics[scale=0.7]{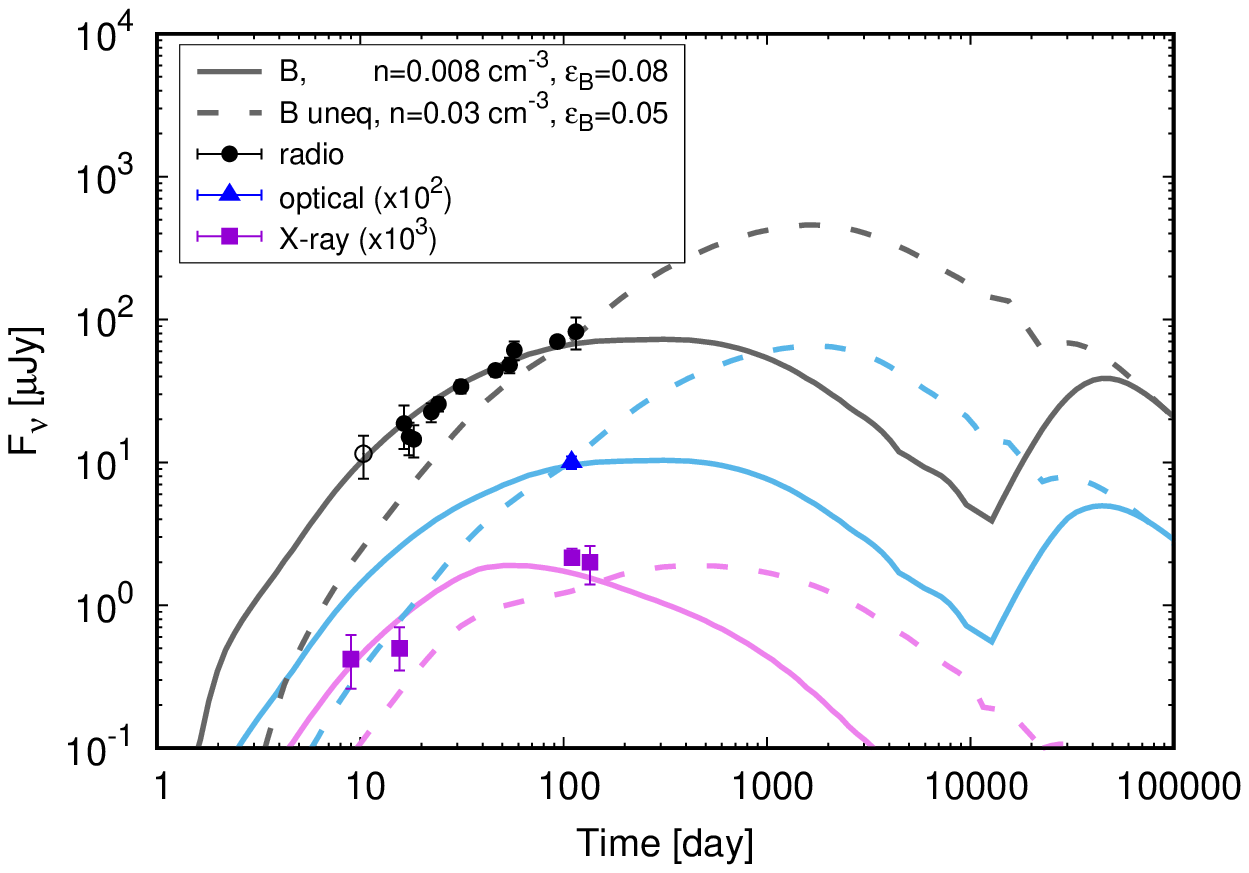}
\caption{
Same as figure \ref{fig:L} but on a longer time scale. 
The radio flare arising from the ejecta component that
produced the uv/optical/nIR macronova signal is also shown.
For this component, we assume a mass of $0.04M_{\odot}$
and a single velocity of $0.2c$ ({\it top}) and $0.1c$ ({\it bottom}),
and the ISM density and microphysics
parameters are the same to the dynamical ejecta component of
each model.
}
\label{fig:L2}
\end{figure}

Considering now the total ejected mass as observed from the  uv/optical/IR macronova, we estimate the 
 peak time and flux density of this slow component  \citep{NP11}:
\begin{eqnarray}
t_{\rm p} & \approx & 10^4\,{\rm day}\,\left(\frac{E}{10^{51}\,{\rm erg}} \right)^{\frac{1}{3}}
\left(\frac{n}{0.01\,{\rm cm^{-3}}} \right)^{-\frac{1}{3}}
\left(\frac{\beta}{0.2} \right)^{-\frac{5}{3}},\\
F_{\nu,{\rm p}} &\approx& 2\,{\rm mJy}\, \epsilon_{B,-1}^{0.775}\epsilon_{e,-1}^{1.1}
\left(\frac{E}{10^{51}\,{\rm erg}} \right) \label{eq:F}
\\
& & ~~~~~~\times \left(\frac{n}{0.01\,{\rm cm^{-3}}} \right)^{0.775}\left(\frac{\beta}{0.2} \right)^{1.75}
\left(\frac{\nu}{3\,{\rm GHz}} \right)^{-0.55}, \nonumber
\end{eqnarray}
where $\epsilon_{B,-1}$ and $\epsilon_{e,-1}$ are normalized by $0.1$ and
we set $p=2.1$. Figure \ref{fig:L2} depicts the flux densities from the
dynamical ejecta and the macronova ejecta.  Here we assume that the macronova
ejecta has a mass of $0.04M_{\odot}$ and a single velocity of $0.2c$ ({\it top} panel) and $0.1c$ ({\it bottom} panel),
and set the ISM density and the same microphysics parameters as those used for the
dynamical ejecta models (see figure \ref{fig:L}). 
The bumps around $10^4$ --$10^5$ days in the light curves
are produced by the macronova component.
The actual shape of
the peak of the light curves is expected to be 
broader due to the velocity structure of the macronova
ejecta. {Note that the cooling frequency of the
late emission is around  the uv/optical bands so that the
estimate of  equation (\ref{eq:F}) is applicable up to the optical band and
the X-ray flux density of the macronova component is much fainter than
that estimated by equation (\ref{eq:F}).}

This late-time emission  is one of the  notable differences between the dynamical ejecta and cocoon/structure jet scenarios.
For the cocoon/structure jet scenario, in which the ISM density
is much lower than that of the dynamical ejecta scenario,
the flux density from the macronova ejecta is expected to be
much fainter $\lesssim 50 \,{\rm \mu Jy}$ and the peak time is much longer, $\gtrsim 100$ yr.

{\it Angular size}:
The velocity of the fast tail that produces the afterglow at $100$ days is $\sim 0.6c$.
Therefore, the angular size of the afterglow is  
$\lesssim 500\,{\rm \mu arcsec} (t/100\,{\rm day})$ for $> 100$ days.
 This is smaller by a factor a  few compared to that expected in the cocoon and structured jet models.
Therefore, these scenarios can be distinguished by future  VLBI observations  (\citealt{Hallinan2017}).

\section{Conclusion}
\label{sec:conclusions}
The fast tail ($\Gamma \beta \sim 1$) of the dynamical ejecta  of binary neutron star mergers, calculated in recent high resolution numerical simulations by \cite{Kiuchi2017}
contains kinetic energy of $10^{47}$ -- $10^{49}$ erg, depending on the neutron star equation 
of state and on the binary masses. Mergers with  a softer neutron star equation of state, which gives
 smaller radius neutron stars, and 
with a mass ratio close to unity eject larger amounts of the fast tail. The fast tail has somewhat isotropic 
shape even for  models in which the bulk of the material is  large aspherical.


The interaction of this fast tail with  the surrounding ISM produces a blast wave whose synchrotron emission \citep{NP11} is  observed as the radio to X-ray signals that followed GW170817. 
We compute this synchrotron emission  arising from  the ejecta profile obtained from the high resolution  numerical  relativity simulations and 
 compare them  with the observed data of the non-thermal radiation that followed  GW170817. 
We find that the multi-frequency observed data  can be
reproduced well for the equal mass binary models with a relatively soft equation of state.
In all cases, an ISM density of $\sim 0.01\,{\rm cm^{-3}}$ is required to obtain 
the observed flux level at $\sim 100$ days after the merger. 
For unequal mass cases, the  velocity gradient of the ejecta profile is steeper and the light curves rise more steeply than the observed one so that nearly equal mass mergers are favored in this scenario.

The dynamical ejecta scenario has three generic predictions:  
\begin{enumerate}
\item The cooling frequency crosses the X-ray band on a time scale of a few months to a year leading to a cooling break in the X-ray light curve.
\item The outer ejecta velocity is $\lesssim 0.6c$ at $t \gtrsim100$ days so that the angular size
of the ejecta will be   $\lesssim 500 \,\mu{\rm arcsec}(t/100\,{\rm day})$.
\item The emission of the sub-relativistic  macronova/kilonova component of a mass of $\sim 0.05M_{\odot}$ and a velocity of
$\sim 0.1c$ will be continuously observable with flux densities of $0.1$--$1$ mJy on time scales of $10^3$ -- $10^5$ days \citep{NP11,piran2013MNRAS,HP2015}. 
\end{enumerate}
These  features will enable us to observationally confirm or rule out this model in the near future.


\acknowledgments
We thank Paolo D'Avanzo, Gregg Hallinan, Raffaella Margutti, Kunal Mooley, David Radice, and Eli Waxman for useful discussions. 
K.H. was supported by Lyman Spitzer Jr. fellowship at Department of Astrophysical sciences at Princeton University.
T.P. was partially supported by an advanced ERC grant TReX and by a grant from the Templeton foundation.
T.P. and E.N. were supported by the I-Core center of excellence of the Israeli Science Foundation. 
Numerical  simulations were performed on K computer at AICS (project
numbers  hp160211, hp170230, hp170313),  on  Cray  XC30  at cfca  of
National  Astronomical  Observatory  of  Japan, FX10 at Information
Technology Center of the University  of  Tokyo,  HOKUSAI  FX100  at
RIKEN,  and  on Cray XC40 at Yukawa Institute for Theoretical Physics,
Kyoto  University. This  work  was  supported  by  Grant-
in-Aid  for  Scientific  Research  (16H02183, 17H06361,  15K05077)  of  JSPS  and
by a  post-K  computer  project  (Priority  issue  No.  9)  of Japanese
MEXT.


\begin{thebibliography}{}
\expandafter\ifx\csname natexlab\endcsname\relax\def\natexlab#1{#1}\fi

\bibitem[{Abbott {et~al.}(2017)}]{GW170817}
Abbott, B.~P., {et~al.} 2017, Phys. Rev. Lett., 119, 161101

\bibitem[{{Abbott} {et~al.}(2017){Abbott}, {Abbott}, {Abbott}, {Acernese},
  {Ackley}, {Adams}, {Adams}, {Addesso}, {Adhikari}, {Adya}, \&
  et~al.}]{Abbott2017ApJ}
{Abbott}, B.~P., {Abbott}, R., {Abbott}, T.~D., {et~al.} 2017, \apjl, 848, L12

\bibitem[{Andreoni {et~al.}(2017)}]{Andreoni2017}
Andreoni, I., {et~al.} 2017, Submitted to: Publ. Astron. Soc. Austral.,
  arXiv:1710.05846

\bibitem[{{Arcavi} {et~al.}(2017){Arcavi}, {Hosseinzadeh}, {Howell}, {McCully},
  {Poznanski}, {Kasen}, {Barnes}, {Zaltzman}, {Vasylyev}, {Maoz}, \&
  {Valenti}}]{Arcavi2017Natur}
{Arcavi}, I., {Hosseinzadeh}, G., {Howell}, D.~A., {et~al.} 2017, \nat, 551, 64

\bibitem[{{Bauswein} {et~al.}(2013){Bauswein}, {Goriely}, \&
  {Janka}}]{bauswein2013ApJ}
{Bauswein}, A., {Goriely}, S., \& {Janka}, H.-T. 2013, \apj, 773, 78

\bibitem[{{C{\^o}t{\'e}} {et~al.}(2017){C{\^o}t{\'e}}, {Fryer}, {Belczynski},
  {Korobkin}, {Chru{\'s}li{\'n}ska}, {Vassh}, {Mumpower}, {Lippuner},
  {Sprouse}, {Surman}, \& {Wollaeger}}]{Cote2017}
{C{\^o}t{\'e}}, B., {Fryer}, C.~L., {Belczynski}, K., {et~al.} 2017, ArXiv
  e-prints, arXiv:1710.05875

\bibitem[{{Coulter} {et~al.}(2017){Coulter}, {Foley}, {Kilpatrick}, {Drout},
  {Piro}, {Shappee}, {Siebert}, {Simon}, {Ulloa}, {Kasen}, {Madore},
  {Murguia-Berthier}, {Pan}, {Prochaska}, {Ramirez-Ruiz}, {Rest}, \&
  {Rojas-Bravo}}]{Swope}
{Coulter}, D.~A., {Foley}, R.~J., {Kilpatrick}, C.~D., {et~al.} 2017, Science,
  358, 1556

\bibitem[{Cowperthwaite {et~al.}(2017)}]{Cowperthwaite2017}
Cowperthwaite, P.~S., {et~al.} 2017, Astrophys. J., 848, L17

\bibitem[{{D'Avanzo} {et~al.}(2018){D'Avanzo}, {Campana}, {Ghisellini},
  {Melandri}, {Bernardini}, {Covino}, {D'Elia}, {Nava}, {Salvaterra},
  {Tagliaferri}, \& {Vergani}}]{D'Avanzo2018}
{D'Avanzo}, P., {Campana}, S., {Ghisellini}, G., {et~al.} 2018, ArXiv e-prints,
  arXiv:1801.06164

\bibitem[{{Dessart} {et~al.}(2009){Dessart}, {Ott}, {Burrows}, {Rosswog}, \&
  {Livne}}]{dessart09}
{Dessart}, L., {Ott}, C.~D., {Burrows}, A., {Rosswog}, S., \& {Livne}, E. 2009,
  ApJ, 690, 1681

\bibitem[{{Dietrich} {et~al.}(2017){Dietrich}, {Ujevic}, {Tichy}, {Bernuzzi},
  \& {Br{\"u}gmann}}]{Dietrich2017}
{Dietrich}, T., {Ujevic}, M., {Tichy}, W., {Bernuzzi}, S., \& {Br{\"u}gmann},
  B. 2017, \prd, 95, 024029

\bibitem[{{Drout} {et~al.}(2017){Drout}, {Piro}, {Shappee}, {Kilpatrick},
  {Simon}, {Contreras}, {Coulter}, {Foley}, {Siebert}, {Morrell}, {Boutsia},
  {Di Mille}, {Holoien}, {Kasen}, {Kollmeier}, {Madore}, {Monson},
  {Murguia-Berthier}, {Pan}, {Prochaska}, {Ramirez-Ruiz}, {Rest}, {Adams},
  {Alatalo}, {Ba{\~n}ados}, {Baughman}, {Beers}, {Bernstein}, {Bitsakis},
  {Campillay}, {Hansen}, {Higgs}, {Ji}, {Maravelias}, {Marshall}, {Bidin},
  {Prieto}, {Rasmussen}, {Rojas-Bravo}, {Strom}, {Ulloa},
  {Vargas-Gonz{\'a}lez}, {Wan}, \& {Whitten}}]{Drout2017}
{Drout}, M.~R., {Piro}, A.~L., {Shappee}, B.~J., {et~al.} 2017, Science, 358,
  1570

\bibitem[{{Eichler} {et~al.}(1989){Eichler}, {Livio}, {Piran}, \&
  {Schramm}}]{eichler1989Nature}
{Eichler}, D., {Livio}, M., {Piran}, T., \& {Schramm}, D.~N. 1989, \nat, 340,
  126

\bibitem[{{Evans} {et~al.}(2017){Evans}, {Cenko}, {Kennea}, {Emery}, {Kuin},
  {Korobkin}, {Wollaeger}, {Fryer}, {Madsen}, {Harrison}, {Xu}, {Nakar},
  {Hotokezaka}, {Lien}, {Campana}, {Oates}, {Troja}, {Breeveld}, {Marshall},
  {Barthelmy}, {Beardmore}, {Burrows}, {Cusumano}, {D{\'}A{\`i}},
  {D{\'}Avanzo}, {D{\'}Elia}, {de Pasquale}, {Even}, {Fontes},
  {Forster}, {Garcia}, {Giommi}, {Grefenstette}, {Gronwall}, {Hartmann},
  {Heida}, {Hungerford}, {Kasliwal}, {Krimm}, {Levan}, {Malesani}, {Melandri},
  {Miyasaka}, {Nousek}, {O{\'}Brien}, {Osborne}, {Pagani}, {Page},
  {Palmer}, {Perri}, {Pike}, {Racusin}, {Rosswog}, {Siegel}, {Sakamoto},
  {Sbarufatti}, {Tagliaferri}, {Tanvir}, \& {Tohuvavohu}}]{Evans2017}
{Evans}, P.~A., {Cenko}, S.~B., {Kennea}, J.~A., {et~al.} 2017, Science, 358,
  1565

\bibitem[{{Fern{\'a}ndez} \& {Metzger}(2013)}]{fernandez2013MNRAS}
{Fern{\'a}ndez}, R., \& {Metzger}, B.~D. 2013, \mnras, 435, 502

\bibitem[{{Fern{\'a}ndez} {et~al.}(2015){Fern{\'a}ndez}, {Quataert}, {Schwab},
  {Kasen}, \& {Rosswog}}]{fernandez2015MNRAS}
{Fern{\'a}ndez}, R., {Quataert}, E., {Schwab}, J., {Kasen}, D., \& {Rosswog},
  S. 2015, \mnras, 449, 390

\bibitem[{{Fujibayashi} {et~al.}(2017){Fujibayashi}, {Kiuchi}, {Nishimura},
  {Sekiguchi}, \& {Shibata}}]{fujibayashi2017}
{Fujibayashi}, S., {Kiuchi}, K., {Nishimura}, N., {Sekiguchi}, Y., \&
  {Shibata}, M. 2017, ArXiv e-prints, arXiv:1711.02093

\bibitem[{{Gottlieb} {et~al.}(2018){Gottlieb}, {Nakar}, \&
  {Piran}}]{Gottlieb2018MNRAS}
{Gottlieb}, O., {Nakar}, E., \& {Piran}, T. 2018, \mnras, 473, 576

\bibitem[{{Gottlieb} {et~al.}(2017){Gottlieb}, {Nakar}, {Piran}, \&
  {Hotokezaka}}]{Gottlieb+17B}
{Gottlieb}, O., {Nakar}, E., {Piran}, T., \& {Hotokezaka}, K. 2017, ArXiv
  e-prints, arXiv:1710.05896

\bibitem[{{Granot} \& {Sari}(2002)}]{GranotSari02}
{Granot}, J., \& {Sari}, R. 2002, \apj, 568, 820

\bibitem[{Haggard {et~al.}(2017)Haggard, Nynka, Ruan, Kalogera, Bradley~Cenko,
  Evans, \& Kennea}]{Haggard2017}
Haggard, D., Nynka, M., Ruan, J.~J., {et~al.} 2017, Astrophys. J., 848, L25

\bibitem[{{Hallinan} {et~al.}(2017){Hallinan}, {Corsi}, {Mooley}, {Hotokezaka},
  {Nakar}, {Kasliwal}, {Kaplan}, {Frail}, {Myers}, {Murphy}, {De}, {Dobie},
  {Allison}, {Bannister}, {Bhalerao}, {Chandra}, {Clarke}, {Giacintucci}, {Ho},
  {Horesh}, {Kassim}, {Kulkarni}, {Lenc}, {Lockman}, {Lynch}, {Nichols},
  {Nissanke}, {Palliyaguru}, {Peters}, {Piran}, {Rana}, {Sadler}, \&
  {Singer}}]{Hallinan2017}
{Hallinan}, G., {Corsi}, A., {Mooley}, K.~P., {et~al.} 2017, Science, 358, 1579

\bibitem[{{Hotokezaka} {et~al.}(2018){Hotokezaka}, {Beniamini}, \&
  {Piran}}]{hotokezaka2018}
{Hotokezaka}, K., {Beniamini}, P., \& {Piran}, T. 2018, ArXiv e-prints,
  arXiv:1801.01141

\bibitem[{{Hotokezaka} {et~al.}(2013){Hotokezaka}, {Kiuchi}, {Kyutoku},
  {Okawa}, {Sekiguchi}, {Shibata}, \& {Taniguchi}}]{hotokezaka13b}
{Hotokezaka}, K., {Kiuchi}, K., {Kyutoku}, K., {et~al.} 2013, Phys. Rev. D, 87,
  024001

\bibitem[{{Hotokezaka} \& {Piran}(2015)}]{HP2015}
{Hotokezaka}, K., \& {Piran}, T. 2015, \mnras, 450, 1430

\bibitem[{{Just} {et~al.}(2015){Just}, {Bauswein}, {Pulpillo}, {Goriely}, \&
  {Janka}}]{just2015MNRAS}
{Just}, O., {Bauswein}, A., {Pulpillo}, R.~A., {Goriely}, S., \& {Janka}, H.-T.
  2015, \mnras, 448, 541

\bibitem[{{Kasen} {et~al.}(2017){Kasen}, {Metzger}, {Barnes}, {Quataert}, \&
  {Ramirez-Ruiz}}]{Kasen2017}
{Kasen}, D., {Metzger}, B., {Barnes}, J., {Quataert}, E., \& {Ramirez-Ruiz}, E.
  2017, \nat, 551, 80

\bibitem[{{Kasliwal} {et~al.}(2017){Kasliwal}, {Nakar}, {Singer}, {Kaplan},
  {Cook}, {Van Sistine}, {Lau}, {Fremling}, {Gottlieb}, {Jencson}, {Adams},
  {Feindt}, {Hotokezaka}, {Ghosh}, {Perley}, {Yu}, {Piran}, {Allison},
  {Anupama}, {Balasubramanian}, {Bannister}, {Bally}, {Barnes}, {Barway},
  {Bellm}, {Bhalerao}, {Bhattacharya}, {Blagorodnova}, {Bloom}, {Brady},
  {Cannella}, {Chatterjee}, {Cenko}, {Cobb}, {Copperwheat}, {Corsi}, {De},
  {Dobie}, {Emery}, {Evans}, {Fox}, {Frail}, {Frohmaier}, {Goobar}, {Hallinan},
  {Harrison}, {Helou}, {Hinderer}, {Ho}, {Horesh}, {Ip}, {Itoh}, {Kasen},
  {Kim}, {Kuin}, {Kupfer}, {Lynch}, {Madsen}, {Mazzali}, {Miller}, {Mooley},
  {Murphy}, {Ngeow}, {Nichols}, {Nissanke}, {Nugent}, {Ofek}, {Qi}, {Quimby},
  {Rosswog}, {Rusu}, {Sadler}, {Schmidt}, {Sollerman}, {Steele}, {Williamson},
  {Xu}, {Yan}, {Yatsu}, {Zhang}, \& {Zhao}}]{Kasliwal2017}
{Kasliwal}, M.~M., {Nakar}, E., {Singer}, L.~P., {et~al.} 2017, Science, 358,
  1559

\bibitem[{Kilpatrick {et~al.}(2017)}]{Kilpatrick2017}
Kilpatrick, C.~D., {et~al.} 2017, arXiv:1710.05434

\bibitem[{{Kiuchi} {et~al.}(2017){Kiuchi}, {Kawaguchi}, {Kyutoku}, {Sekiguchi},
  {Shibata}, \& {Taniguchi}}]{Kiuchi2017}
{Kiuchi}, K., {Kawaguchi}, K., {Kyutoku}, K., {et~al.} 2017, \prd, 96, 084060

\bibitem[{{Kyutoku} {et~al.}(2014){Kyutoku}, {Ioka}, \&
  {Shibata}}]{Kyutoku2014}
{Kyutoku}, K., {Ioka}, K., \& {Shibata}, M. 2014, \mnras, 437, L6

\bibitem[{{Lattimer} \& {Schramm}(1974)}]{Lattimer1974}
{Lattimer}, J.~M., \& {Schramm}, D.~N. 1974, \apjl, 192, L145

\bibitem[{{Lazzati} {et~al.}(2017){Lazzati}, {Perna}, {Morsony},
  {L{\'o}pez-C{\'a}mara}, {Cantiello}, {Ciolfi}, {giacomazzo}, \&
  {Workman}}]{Lazzati2017}
{Lazzati}, D., {Perna}, R., {Morsony}, B.~J., {et~al.} 2017, ArXiv e-prints,
  arXiv:1712.03237

\bibitem[{{Lipunov} {et~al.}(2017){Lipunov}, {Gorbovskoy}, {Kornilov},
  {.~Tyurina}, {Balanutsa}, {Kuznetsov}, {Vlasenko}, {Kuvshinov}, {Gorbunov},
  {Buckley}, {Krylov}, {Podesta}, {Lopez}, {Podesta}, {Levato}, {Saffe},
  {Mallamachi}, {Potter}, {Budnev}, {Gress}, {Ishmuhametova}, {Vladimirov},
  {Zimnukhov}, {Yurkov}, {Sergienko}, {Gabovich}, {Rebolo}, {Serra-Ricart},
  {Israelyan}, {Chazov}, {Wang}, {Tlatov}, \& {Panchenko}}]{Lipunov2017}
{Lipunov}, V.~M., {Gorbovskoy}, E., {Kornilov}, V.~G., {et~al.} 2017, \apjl,
  850, L1

\bibitem[{{Lyman} {et~al.}(2018){Lyman}, {Lamb}, {Levan}, {Mandel}, {Tanvir},
  {Kobayashi}, {Gompertz}, {Hjorth}, {Fruchter}, {Kangas}, {Steeghs}, {Steele},
  {Cano}, {Copperwheat}, {Evans}, {Fynbo}, {Gall}, {Im}, {Izzo}, {Jakobsson},
  {Milvang-Jensen}, {O'Brien}, {Osborne}, {Palazzi}, {Perley}, {Pian},
  {Rosswog}, {Rowlinson}, {Schulze}, {Stanway}, {Sutton}, {Th{\"o}ne}, {de
  Ugarte Postigo}, {Watson}, {Wiersema}, \& {Wijers}}]{Lyman2018}
{Lyman}, J.~D., {Lamb}, G.~P., {Levan}, A.~J., {et~al.} 2018, ArXiv e-prints,
  arXiv:1801.02669

\bibitem[{{Margutti} {et~al.}(2017){Margutti}, {Berger}, {Fong}, {Guidorzi},
  {Alexander}, {Metzger}, {Blanchard}, {Cowperthwaite}, {Chornock},
  {Eftekhari}, {Nicholl}, {Villar}, {Williams}, {Annis}, {Brown}, {Chen},
  {Doctor}, {Frieman}, {Holz}, {Sako}, \& {Soares-Santos}}]{Margutti2017}
{Margutti}, R., {Berger}, E., {Fong}, W., {et~al.} 2017, \apjl, 848, L20

\bibitem[{{Margutti} {et~al.}(2018){Margutti}, {Alexander}, {Xie}, {Sironi},
  {Metzger}, {Kathirgamaraju}, {Fong}, {Blanchard}, {Berger}, {MacFadyen},
  {Giannios}, {Guidorzi}, {Hajela}, {Chornock}, {Cowperthwaite}, {Eftekhari},
  {Nicholl}, {Villar}, {Williams}, \& {Zrake}}]{Margutti2018}
{Margutti}, R., {Alexander}, K.~D., {Xie}, X., {et~al.} 2018, ArXiv e-prints,
  arXiv:1801.03531
  
  
\bibitem[{{Metzger} {et~al.}(2018){Metzger}, {Thompson}, \&
  {Quataert}}]{Metzger2018}
{Metzger}, B.~D., {Thompson}, T.~A., \& {Quataert}, E. 2018, ArXiv e-prints,
  arXiv:1801.04286



\bibitem[{{Metzger} {et~al.}(2015){Metzger}, {Bauswein}, {Goriely}, \&
  {Kasen}}]{Metzger2015}
{Metzger}, B.~D., {Bauswein}, A., {Goriely}, S., \& {Kasen}, D. 2015, \mnras,
  446, 1115

\bibitem[{{Metzger} \& {Fern{\'a}ndez}(2014)}]{metzger2014MNRAS}
{Metzger}, B.~D., \& {Fern{\'a}ndez}, R. 2014, \mnras, 441, 3444

\bibitem[{{Mooley} {et~al.}(2018){Mooley}, {Nakar}, {Hotokezaka}, {Hallinan},
  {Corsi}, {Frail}, {Horesh}, {Murphy}, {Lenc}, {Kaplan}, {de}, {Dobie},
  {Chandra}, {Deller}, {Gottlieb}, {Kasliwal}, {Kulkarni}, {Myers}, {Nissanke},
  {Piran}, {Lynch}, {Bhalerao}, {Bourke}, {Bannister}, \&
  {Singer}}]{Mooley2017}
{Mooley}, K.~P., {Nakar}, E., {Hotokezaka}, K., {et~al.} 2018, \nat, 554, 207

\bibitem[{{Nakar} \& {Piran}(2011)}]{NP11}
{Nakar}, E., \& {Piran}, T. 2011, \nat, 478, 82

\bibitem[{{Nakar} \& {Piran}(2018)}]{Nakar2018}
---. 2018, ArXiv e-prints, arXiv:1801.09712

\bibitem[{{Perego} {et~al.}(2017){Perego}, {Radice}, \&
  {Bernuzzi}}]{Perego2017}
{Perego}, A., {Radice}, D., \& {Bernuzzi}, S. 2017, \apjl, 850, L37

\bibitem[{{Perego} {et~al.}(2014){Perego}, {Rosswog}, {Cabez{\'o}n},
  {Korobkin}, {K{\"a}ppeli}, {Arcones}, \&
  {Liebend{\"o}rfer}}]{perego2014MNRAS}
{Perego}, A., {Rosswog}, S., {Cabez{\'o}n}, R.~M., {et~al.} 2014, \mnras, 443,
  3134

\bibitem[{{Pian} {et~al.}(2017){Pian}, {D'Avanzo}, {Benetti}, {Branchesi},
  {Brocato}, {Campana}, {Cappellaro}, {Covino}, {D'Elia}, {Fynbo}, {Getman},
  {Ghirlanda}, {Ghisellini}, {Grado}, {Greco}, {Hjorth}, {Kouveliotou},
  {Levan}, {Limatola}, {Malesani}, {Mazzali}, {Melandri}, {M{\o}ller},
  {Nicastro}, {Palazzi}, {Piranomonte}, {Rossi}, {Salafia}, {Selsing},
  {Stratta}, {Tanaka}, {Tanvir}, {Tomasella}, {Watson}, {Yang}, {Amati},
  {Antonelli}, {Ascenzi}, {Bernardini}, {Bo{\"e}r}, {Bufano}, {Bulgarelli},
  {Capaccioli}, {Casella}, {Castro-Tirado}, {Chassande-Mottin}, {Ciolfi},
  {Copperwheat}, {Dadina}, {De Cesare}, {di Paola}, {Fan}, {Gendre},
  {Giuffrida}, {Giunta}, {Hunt}, {Israel}, {Jin}, {Kasliwal}, {Klose}, {Lisi},
  {Longo}, {Maiorano}, {Mapelli}, {Masetti}, {Nava}, {Patricelli}, {Perley},
  {Pescalli}, {Piran}, {Possenti}, {Pulone}, {Razzano}, {Salvaterra},
  {Schipani}, {Spera}, {Stamerra}, {Stella}, {Tagliaferri}, {Testa}, {Troja},
  {Turatto}, {Vergani}, \& {Vergani}}]{Pian2017}
{Pian}, E., {D'Avanzo}, P., {Benetti}, S., {et~al.} 2017, \nat, 551, 67

\bibitem[{{Piran} {et~al.}(2013){Piran}, {Nakar}, \&
  {Rosswog}}]{piran2013MNRAS}
{Piran}, T., {Nakar}, E., \& {Rosswog}, S. 2013, \mnras, 430, 2121

\bibitem[{{Radice} {et~al.}(2016){Radice}, {Galeazzi}, {Lippuner}, {Roberts},
  {Ott}, \& {Rezzolla}}]{radice2016MNRAS}
{Radice}, D., {Galeazzi}, F., {Lippuner}, J., {et~al.} 2016, \mnras, 460, 3255

\bibitem[{{Rosswog} {et~al.}(2017){Rosswog}, {Sollerman}, {Feindt}, {Goobar},
  {Korobkin}, {Fremling}, \& {Kasliwal}}]{rosswog2017}
{Rosswog}, S., {Sollerman}, J., {Feindt}, U., {et~al.} 2017, ArXiv e-prints,
  arXiv:1710.05445

\bibitem[{{Ruan} {et~al.}(2017){Ruan}, {Nynka}, {Haggard}, {Kalogera}, \&
  {Evans}}]{Ruan2017}
{Ruan}, J.~J., {Nynka}, M., {Haggard}, D., {Kalogera}, V., \& {Evans}, P. 2017,
  ArXiv e-prints, arXiv:1712.02809

\bibitem[{{Sari} {et~al.}(1998){Sari}, {Piran}, \& {Narayan}}]{SPN98}
{Sari}, R., {Piran}, T., \& {Narayan}, R. 1998, \apjl, 497, L17+

\bibitem[{{Sekiguchi} {et~al.}(2015){Sekiguchi}, {Kiuchi}, {Kyutoku}, \&
  {Shibata}}]{sekiguchi2015PRD}
{Sekiguchi}, Y., {Kiuchi}, K., {Kyutoku}, K., \& {Shibata}, M. 2015, \prd, 91,
  064059

\bibitem[{{Shibata} {et~al.}(2017){Shibata}, {Fujibayashi}, {Hotokezaka},
  {Kiuchi}, {Kyutoku}, {Sekiguchi}, \& {Tanaka}}]{Shibata2017}
{Shibata}, M., {Fujibayashi}, S., {Hotokezaka}, K., {et~al.} 2017, \prd, 96,
  123012

\bibitem[{{Siegel} \& {Metzger}(2017)}]{Siegel2017}
{Siegel}, D.~M., \& {Metzger}, B.~D. 2017, ArXiv e-prints, arXiv:1705.05473

\bibitem[{{Smartt} {et~al.}(2017){Smartt}, {Chen}, {Jerkstrand}, {Coughlin},
  {Kankare}, {Sim}, {Fraser}, {Inserra}, {Maguire}, {Chambers}, {Huber},
  {Kr{\"u}hler}, {Leloudas}, {Magee}, {Shingles}, {Smith}, {Young}, {Tonry},
  {Kotak}, {Gal-Yam}, {Lyman}, {Homan}, {Agliozzo}, {Anderson}, {Angus},
  {Ashall}, {Barbarino}, {Bauer}, {Berton}, {Botticella}, {Bulla}, {Bulger},
  {Cannizzaro}, {Cano}, {Cartier}, {Cikota}, {Clark}, {De Cia}, {Della Valle},
  {Denneau}, {Dennefeld}, {Dessart}, {Dimitriadis}, {Elias-Rosa}, {Firth},
  {Flewelling}, {Fl{\"o}rs}, {Franckowiak}, {Frohmaier}, {Galbany},
  {Gonz{\'a}lez-Gait{\'a}n}, {Greiner}, {Gromadzki}, {Guelbenzu},
  {Guti{\'e}rrez}, {Hamanowicz}, {Hanlon}, {Harmanen}, {Heintz}, {Heinze},
  {Hernandez}, {Hodgkin}, {Hook}, {Izzo}, {James}, {Jonker}, {Kerzendorf},
  {Klose}, {Kostrzewa-Rutkowska}, {Kowalski}, {Kromer}, {Kuncarayakti},
  {Lawrence}, {Lowe}, {Magnier}, {Manulis}, {Martin-Carrillo}, {Mattila},
  {McBrien}, {M{\"u}ller}, {Nordin}, {O'Neill}, {Onori}, {Palmerio},
  {Pastorello}, {Patat}, {Pignata}, {Podsiadlowski}, {Pumo}, {Prentice}, {Rau},
  {Razza}, {Rest}, {Reynolds}, {Roy}, {Ruiter}, {Rybicki}, {Salmon}, {Schady},
  {Schultz}, {Schweyer}, {Seitenzahl}, {Smith}, {Sollerman}, {Stalder},
  {Stubbs}, {Sullivan}, {Szegedi}, {Taddia}, {Taubenberger}, {Terreran}, {van
  Soelen}, {Vos}, {Wainscoat}, {Walton}, {Waters}, {Weiland}, {Willman},
  {Wiseman}, {Wright}, {Wyrzykowski}, \& {Yaron}}]{Smartt2017}
{Smartt}, S.~J., {Chen}, T.-W., {Jerkstrand}, A., {et~al.} 2017, \nat, 551, 75

\bibitem[{Tanaka {et~al.}(2017)}]{Tanaka2017}
Tanaka, M., {et~al.} 2017, Publ. Astron. Soc. Jap., arXiv:1710.05850

\bibitem[{Tanvir {et~al.}(2017)}]{Tanvir2017}
Tanvir, N.~R., {et~al.} 2017, Astrophys. J., 848, L27

\bibitem[{{Troja} {et~al.}(2017){Troja}, {Piro}, {van Eerten}, {Wollaeger},
  {Im}, {Fox}, {Butler}, {Cenko}, {Sakamoto}, {Fryer}, {Ricci}, {Lien}, {Ryan},
  {Korobkin}, {Lee}, {Burgess}, {Lee}, {Watson}, {Choi}, {Covino}, {D'Avanzo},
  {Fontes}, {Gonz{\'a}lez}, {Khandrika}, {Kim}, {Kim}, {Lee}, {Lee}, {Kutyrev},
  {Lim}, {S{\'a}nchez-Ram{\'{\i}}rez}, {Veilleux}, {Wieringa}, \&
  {Yoon}}]{Troja2017}
{Troja}, E., {Piro}, L., {van Eerten}, H., {et~al.} 2017, \nat, 551, 71

\bibitem[{{Troja} {et~al.}(2018){Troja}, {Piro}, {Ryan}, {van Eerten}, {Ricci},
  {Wieringa}, {Lotti}, {Sakamoto}, \& {Cenko}}]{Troja2018}
{Troja}, E., {Piro}, L., {Ryan}, G., {et~al.} 2018, ArXiv e-prints,
  arXiv:1801.06516

\bibitem[{Utsumi {et~al.}(2017)}]{Utsumi2017}
Utsumi, Y., {et~al.} 2017, Publ. Astron. Soc. Japan, arXiv:1710.05848

\bibitem[{Valenti {et~al.}(2017)Valenti, Sand, Yang, Cappellaro, Tartaglia,
  Corsi, Jha, Reichart, Haislip, \& Kouprianov}]{Valenti2017}
Valenti, S., Sand, D.~J., Yang, S., {et~al.} 2017, Astrophys. J., 848, L24

\bibitem[{{Villar} {et~al.}(2017){Villar}, {Guillochon}, {Berger}, {Metzger},
  {Cowperthwaite}, {Nicholl}, {Alexander}, {Blanchard}, {Chornock},
  {Eftekhari}, {Fong}, {Margutti}, \& {Williams}}]{Villar2017}
{Villar}, V.~A., {Guillochon}, J., {Berger}, E., {et~al.} 2017, \apjl, 851, L21


\bibitem[{{Waxman} {et~al.}(2017){Waxman}, {Ofek}, {Kushnir}, \&
  {Gal-Yam}}]{Waxman2017}
{Waxman}, E., {Ofek}, E., {Kushnir}, D., \& {Gal-Yam}, A. 2017, ArXiv e-prints,
  arXiv:1711.09638

\end{thebibliography}
\end{document}